\documentclass{aastex63}
\usepackage{mathrsfs}

\graphicspath{{./}{figures/}}

\submitjournal{ApJ}

\shorttitle{NGC 6256}
\shortauthors{M. Cadelano et al.}

\begin{document}

\title{Digging for relics of the past: the ancient and obscured bulge globular cluster NGC~6256}


\author[0000-0002-5038-3914]{Mario Cadelano}
\affil{Dipartimento di Fisica e Astronomia, Università di Bologna, Via Gobetti 93/2 I-40129 Bologna, Italy}
\affil{INAF-Osservatorio di Astrofisica e Scienze dello Spazio di Bologna, Via Gobetti 93/3 I-40129 Bologna, Italy}

\author[0000-0003-4746-6003]{Sara Saracino}
\affil{Astrophysics Research Institute, Liverpool John Moores University, 146 Brownlow Hill, Liverpool L3 5RF, UK}
\affil{INAF-Osservatorio di Astrofisica e Scienze dello Spazio di Bologna, Via Gobetti 93/3 I-40129 Bologna, Italy}

\author[0000-0003-4237-4601]{Emanuele Dalessandro}
\affil{INAF-Osservatorio di Astrofisica e Scienze dello Spazio di Bologna, Via Gobetti 93/3 I-40129 Bologna, Italy}

\author[0000-0002-2165-8528]{Francesco R. Ferraro}
\affil{Dipartimento di Fisica e Astronomia, Università di Bologna, Via Gobetti 93/2 I-40129 Bologna, Italy}
\affil{INAF-Osservatorio di Astrofisica e Scienze dello Spazio di Bologna, Via Gobetti 93/3 I-40129 Bologna, Italy}

\author[0000-0001-5613-4938]{Barbara Lanzoni}
\affil{Dipartimento di Fisica e Astronomia, Università di Bologna, Via Gobetti 93/2 I-40129 Bologna, Italy}
\affil{INAF-Osservatorio di Astrofisica e Scienze dello Spazio di Bologna, Via Gobetti 93/3 I-40129 Bologna, Italy}

\author[0000-0001-8892-4301]{Davide Massari}
\affil{Dipartimento di Fisica e Astronomia, Università di Bologna, Via Gobetti 93/2 I-40129 Bologna, Italy}
\affil{INAF-Osservatorio di Astrofisica e Scienze dello Spazio di Bologna, Via Gobetti 93/3 I-40129 Bologna, Italy}

\author[0000-0002-7104-2107]{Cristina Pallanca}
\affil{Dipartimento di Fisica e Astronomia, Università di Bologna, Via Gobetti 93/2 I-40129 Bologna, Italy}
\affil{INAF-Osservatorio di Astrofisica e Scienze dello Spazio di Bologna, Via Gobetti 93/3 I-40129 Bologna, Italy}

\author[0000-0002-2744-1928]{Maurizio Salaris}
\affil{Astrophysics Research Institute, Liverpool John Moores University, 146 Brownlow Hill, Liverpool L3 5RF, UK}

\begin{abstract}
We used a set of moderately-deep and high-resolution optical observations obtained with the {\it Hubble Space Telescope} to investigate the properties of the stellar population in the heavily obscured bulge globular cluster NGC~6256. The analysis of the color-magnitude diagram revealed a stellar population with an extended blue horizontal branch and severely affected by differential reddening, which was corrected taking into account color excess variations up to  $\delta E(B-V)\sim0.51$. We implemented a Monte Carlo Markov Chain technique to perform the isochrone fitting of the observed color-magnitude diagram in order to derive the stellar age, the cluster distance and the average color excess in the cluster direction. Using different set of isochrones we found that NGC~6256 is characterised by a very old stellar age around $13.0$ Gyr, with a typical uncertainty of $\sim0.5$ Gyr. We also found an average color excess $E(B-V)=1.19$ and a distance from the Sun of 6.8 kpc. We then derived the cluster gravitational center and measured its absolute proper motion using the Gaia-DR2 catalog. All this was used to back-integrate the cluster orbit in a Galaxy-like potential and measure its integrals of motion. It turned out that NGC~6256 is currently in a low-eccentricity orbit entirely confined within the bulge and its integrals of motion are fully compatible with a cluster purely belonging to the Galaxy native globular cluster population. All these pieces of evidence suggest that NGC~6256 is an extremely old relic of the past history of the Galaxy, formed during the very first stages of its assembly.

\end{abstract}

\keywords{globular cluster: individual (NGC 6256) - technique: photometric}

\section{Introduction}
\label{sec:intro}
Bulge globular clusters (GCs) are ideal tools to trace the properties
(kinematics, chemistry and age) of the stellar populations located in
the inner regions of the Galaxy. However, their observation is highly
challenging, not only because they are situated in a distant and very
crowded region of the Milky Way, but also because they are usually
affected by severe and differential extinction, due to the large and
patchy amount of interposed interstellar dust. This is the reason why
most of the bulge GCs are still poorly studied.
The current picture is that a handful of bulge GCs is characterized by
a relatively low metal content ([Fe/H]$<-1.0$) with respect to the
metallicities typically observed for the majority of both GCs and
field stars residing in the bulge \citep[e.g.,][]{bica16}.  This value
of [Fe/H], together with the significant $\alpha$ enhancement
([$\alpha/$Fe]$\geq 0.3$) measured in these systems, suggests that
they have been generated through an early and fast star formation
burst during the initial stages of the Galaxy assembly \citep[see,
  e.g.,][]{cescutti08}. Interestingly, the first age measurements of
some of these relatively metal-poor systems suggested that they are
indeed very old, with ages around 13 Gyr \citep[see, e.g., the cases
  of NGC 6522, HP1 and Djorgovski
  2;][]{kerber2018,kerber2019,ortolani19b}. More in general, improved
age estimates of bulge GCs are tentatively suggesting the presence of
a correlation between the age and the metallicity of these systems,
with the more metal-poor clusters being older than the more metal-rich
ones \citep[see Figure 16 in][]{saracino19}. A deep investigation of
these systems is thus crucial to constrain the slope of the presumed
age-[Fe/H] relation, which is still completely unknown, but could
bring precious information on the bulge formation processes.

This work is part of an ongoing large program aimed at characterizing
the stellar populations of highly extincted stellar systems orbiting
within the Galactic bulge, which led us to the discovery of the surprising properties of the stellar system Terzan~5 \citep[see][]{ferraro09, ferraro15,
  ferraro16, lanzoni10,origlia11, origlia13,origlia19,massari12, massari14a, massari14b,cadelano18} and other intriguing clusters \citep{saracino15,
  saracino16, saracino19,pallanca19}. This paper focused on NGC~6256.  So far,
only a handful of photometric and spectroscopic works have attempted
to constrain its properties, and a reliable and comprehensive
characterization of this system is thus still lacking. One of the
first photometric studies showing the optical color-magnitude diagram
(CMD) of the cluster was presented by \citet{alcaino83}, who suggested
that NGC~6256 is a metal-rich cluster similar to 47~Tucanae, located
at 11 kpc from the Sun and with a large color excess of
$E(B-V)=0.8$. Different results were then found by \citet{webbink85},
who estimated a much lower metallicity for this system
([Fe/H]$\approx-1.56$), a slightly smaller distance ($d=9.1$ kpc), and
a color-excess $E(B-V)=0.88$. Deeper optical studies
\citep{ortolani99} revised again the situation, suggesting that NGC~6256 presents a blue horizontal branch, an intermediate metallicity
([Fe/H]$\approx-1.3$), a distance of just 6.4 kpc, and an even larger
color excess, $E(B-V)=1.1$. The first high resolution study of the
system, performed through Hubble Space Telescope (HST) observations by \citet{piotto02},
revealed very scattered red giant branch and horizontal branch stars,
hinting to the presence of significant variations of the color excess
even within the small field of view covered by the observations.
Finally, the ground-based near-IR photometric investigation performed
by \citet{valenti07} with the ESO-NTT found [Fe/H]$=-1.63$, a distance
modulus $(m-M)_0=14.79$ (corresponding to 9.1 kpc) and $E(B-V)=1.2$,
while the compilation by \citet{harris10} quotes [Fe/H]$=-1.02$,
$d=10.3$ kpc, and $E(B-V)=1.09$. A recent spectroscopic analysis of
ten giant stars \citep[][but see also \citealt{stephens04}]{vasquez18}
revealed that the stellar population of NGC~6256 is characterized by
[Fe/H]$\approx-1.61$, with a possible intrinsic dispersion of 0.2 dex, implying that NGC~6256 is one of the metal-poorest GC currently located within the Galactic bulge.

These results clearly demonstrate that a detailed and reliable
physical characterization of this cluster is still lacking. Indeed,
even the most recent Gaia DR2 data \citep{gaia2016} are useless to this
aim, because of the large reddening and crowding conditions in the
direction of the system, and they can just provide the proper motion
of the brightest stars \citep{baumgardt2019}. The goal of this study
is thus to obtain a coherent view of the properties of the stellar
population in the inner regions of NGC~6256. To this end, we make use
of a set of moderately-deep and high-resolution images acquired with
the HST.  In Section \ref{sec:data}, the data-set and the data
reduction procedure are described; in Section \ref{sec:dr} we discuss
the differential reddening correction; in Section \ref{sec:params} and
\ref{sec:pm_orbit} we present the determination of the cluster stellar
population properties and of its orbit within the Galaxy. Finally, in
Section \ref{sec:conc}, we summarize the results and draw our
conclusions.

\section{Observations and Data analysis}
\label{sec:data}
This work is based on a set of high-resolution optical observations
obtained with the Wide Field Camera 3 (WFC3) on board the HST, under
GO 11628 (PI: Noyola). It consists of 3 images acquired with the F555W
filter and exposure time of 360 s, and 3 images in the F814W filter
with exposure time of 100 s. The photometric analysis was performed
with {\rm DAOPHOT IV} \citep{stetson87} on the dark, bias, flat and
charge transfer efficiency corrected ``flc'' images \citep[see e.g.][]{cadelano17b,cadelano19}. As a first step,
about 200 stars were selected in each image in order to model the
point spread function (PSF), whose full width at half maximum was set
to 1.5 pixels ($\sim0.06\arcsec$) and sampled within a radius of 10
pixels ($\sim0.4\arcsec$). The PSF models were chosen on the basis of
a $\chi^2$ statistic and, in every image, the best-fit was provided by
a Moffat function \citep{moffat69}. These models were finally applied
to all the sources detected at more than $5\sigma$ from the background
level in each image. Then, we built a master catalog with stars
detected in at least 2 of the available images per filter. At the
corresponding positions of these stars, the photometric fit was forced
in all the other frames by using {\rm DAOPHOT/ALLFRAME}
\citep{stetson94}. Finally, for each star we homogenized the
magnitudes estimated in different images, and their weighted mean and
standard deviation were adopted as the star magnitudes and their
related photometric errors.  Instrumental magnitudes have been
reported to the VEGAMAG system by using the zero-points values for an
aperture of 10 pixels, as quoted in the WFC3
website':\footnote{\url{http://www.stsci.edu/hst/instrumentation/wfc3/data-analysis/photometric-calibration/uvis-photometric-calibration};
  see also \citealp{deustua16}} $ZP_{\rm F555W1}=25.735$, $ZP_{\rm
  F555W2}=25.720$ for the F555W magnitudes of stars observed in chip 1 and in chip 2 of the camera, respectively, and $ZP_{\rm
  F814W1}=24.598$ and $ZP_{\rm F814W2}=24.574$ for the corresponding F814W magnitudes. Finally, we applied appropriate aperture
corrections, evaluated independently for each chip and filter at a
radius of 10 pixels from the stellar centers. The aperture corrections
are of the order of $-0.03$ mags for both the chips in the F555W
observations, and around $-0.01$ mags for both the chips of the F814W
images.

The instrumental positions have been corrected for geometric
distortions following the procedure described by
\citet{bellini11}. They have been then reported to the absolute
coordinate system by cross-correlation with the Gaia DR2 publicly available
catalog \citep{gaia2018}.

\begin{figure}[h] 
\centering
\includegraphics[scale=0.6]{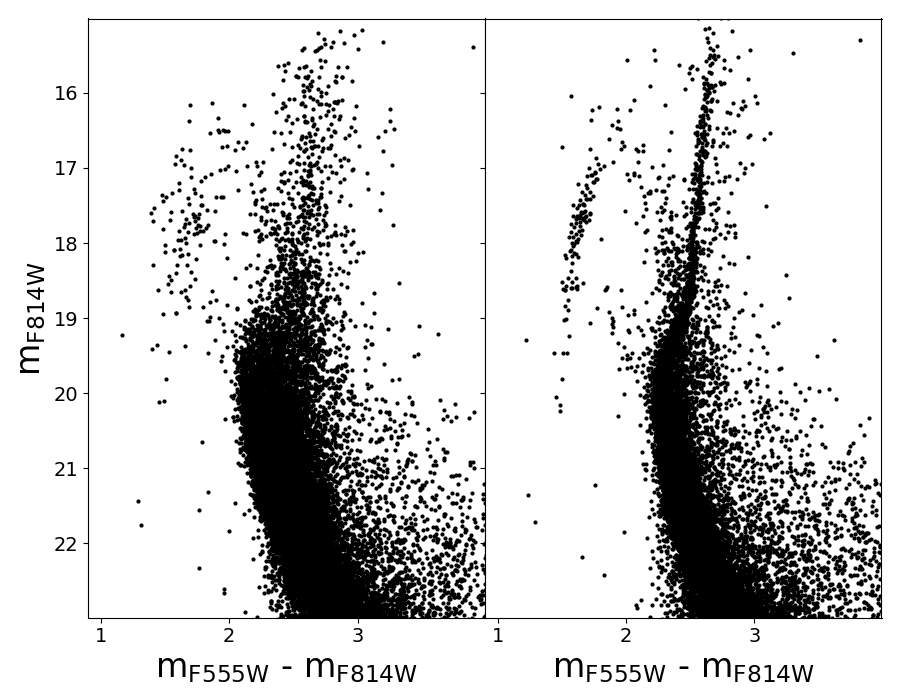}
\caption{{\it Left panel:} Optical CMD of NGC 6256 obtained from the
  HST data-set used in this work. {\it Right panel:} same as in the
  left panel, but after the correction for differential reddening (see Section~\ref{sec:dr}).}
\label{fig:cmd}
\end{figure}

The resulting CMD of the cluster is reported in the left panel of
Figure \ref{fig:cmd}, where it can be appreciated an important
broadening of all the evolutionary sequences due to severe
differential reddening affecting the whole field of view
($\sim160\arcsec \times160\arcsec$).

\section{Differential reddening correction}
\label{sec:dr}
\begin{figure}[t] 
\centering
\includegraphics[scale=0.6]{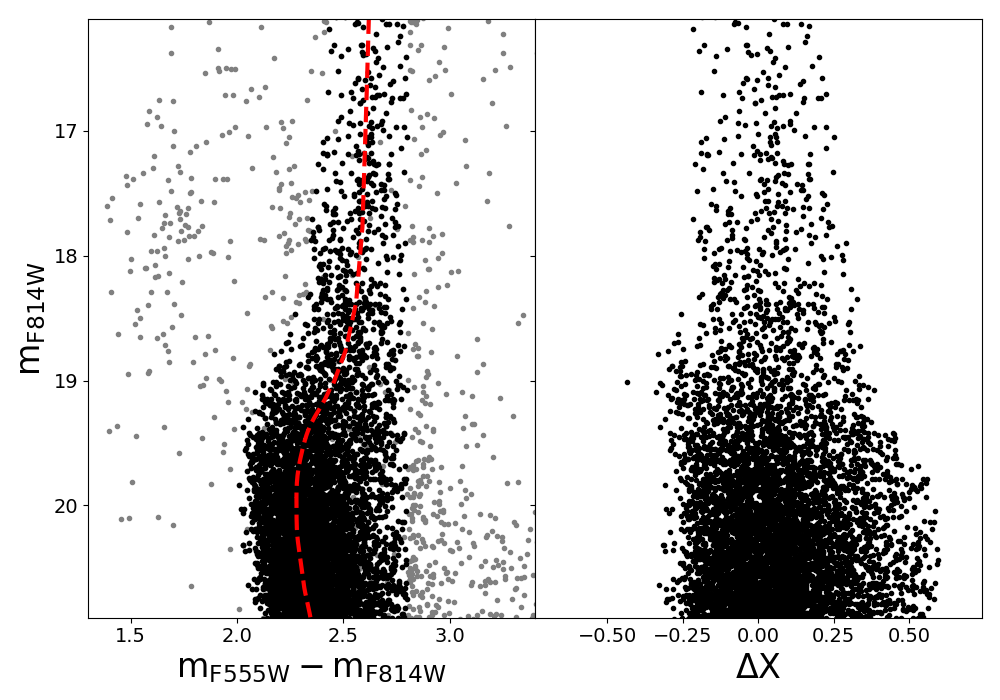}
\caption{{\it Left panel:} Zoomed view of the cluster CMD (the same as
  the one plotted in the left panel of Figure \ref{fig:cmd}). Black points are the stars roughly selected as cluster members and used to
  determine the cluster MRL, which is shown as a dashed red curve.
  {\it Right panel:} stellar magnitudes as a function of their distance  from  the MRL ($\Delta X$)  for the reference sample used for the differential reddening determination (black points on the left panel). }
\label{fig:cmddiff}
\end{figure}

For a proper derivation of the cluster population properties, it is
first necessary to correct the stellar magnitudes for the effects of
the strong differential reddening affecting the field of view. To do
this, we used an approach similar to that described in \citet{dalessandro18} and
\citet[][see also \citealp{milone12}]{saracino19}. First of all, we created the cluster mean ridge
line (MRL). We roughly selected a sample of likely cluster member
stars along the main-sequence, sub-giant and red giant branches based
on their observed position in the CMD. The resulting sample is plotted with black
points in the left panel of Figure \ref{fig:cmddiff}. To minimize the
contamination from field stars, we further selected only the stars located
within the cluster half-light radius  ($\sim50\arcsec$) quoted in \citet{harris10}.  {We then iteratively divided the CMD in magnitude bins of different widths, ranging from 0.15 to 0.5 mags in steps of 0.01 mags. During each iteration, at a fixed bin width, we evaluated the sigma-clipped mean color of the selected sample within each bin and determined a MRL by connecting all these values together. Finally, at the end of the iterations, we averaged all these MRLs to obtain the final MRL shown in the left panel of Figure \ref{fig:cmddiff}.}

For the stars belonging to the sample of likely cluster members in the
magnitude range $16<m_{\rm F814W}<21$, and independently of their
distance from the center, we computed their distance  ($\Delta X$) from the MRL along the reddening vector, defined using the extinction coefficients  $R_{\rm F555W}=3.207$ and $R_{\rm F814W}=1.842$, appropriate for turn-off stars ($T_{\rm
  eff}\sim6000$ K)\footnote{Please note that the extinction coefficients depend on the effective temperature of the stars (see Section~\ref{sec:iso}). By neglecting such a dependence we are introducing a negligible systematic error on the corrected magnitudes. Indeed, such a systematic error is smaller than the photometric errors and significantly smaller than the typical uncertainties on the differential reddening corrections that we are going to derive. See, e.g., \citet{pallanca19}.}  and obtained from  \citet{cardelli89} and \citet{girardi02}.

The resulting distribution of $\Delta
X$ as a function of the stellar magnitude is shown in the right panel of Figure \ref{fig:cmddiff}. This reference sample was used to assign a value of $\Delta X$ to \emph{all} the sources in our photometric catalog, as follows. For each source, $\Delta X$ and its uncertainty were determined as the $\sigma$-clipped median and standard deviation of the $\Delta X$ values measured for the $n$ closest reference stars. The resulting values of $\Delta X$ (and related uncertainties) can be transformed into the local differential reddening $\delta E(B-V)$ and used to correct the observed stellar magnitudes by using the following equation:
\begin{equation}
\delta E(B-V ) = \frac{\Delta X}{ \sqrt{R_{\rm F555W}^2 -2R_{\rm F555W}R_{\rm F814W} + 2R_{\rm F814W}^2  } },
\end{equation}This procedure was iterated three times. Initially, $\Delta X$ was measured for each star using the 30 closest reference stars, then using the 25 and finally the 20 closest stars. These numbers have been chosen as a compromise between having enough statistic and achieving good enough spatial resolution in the final reddening map. The procedure was stopped at the third iteration, as a fourth one would introduce magnitude corrections negligible with respect to the photometric errors.

\begin{figure}[h] 
\centering
\includegraphics[scale=0.6]{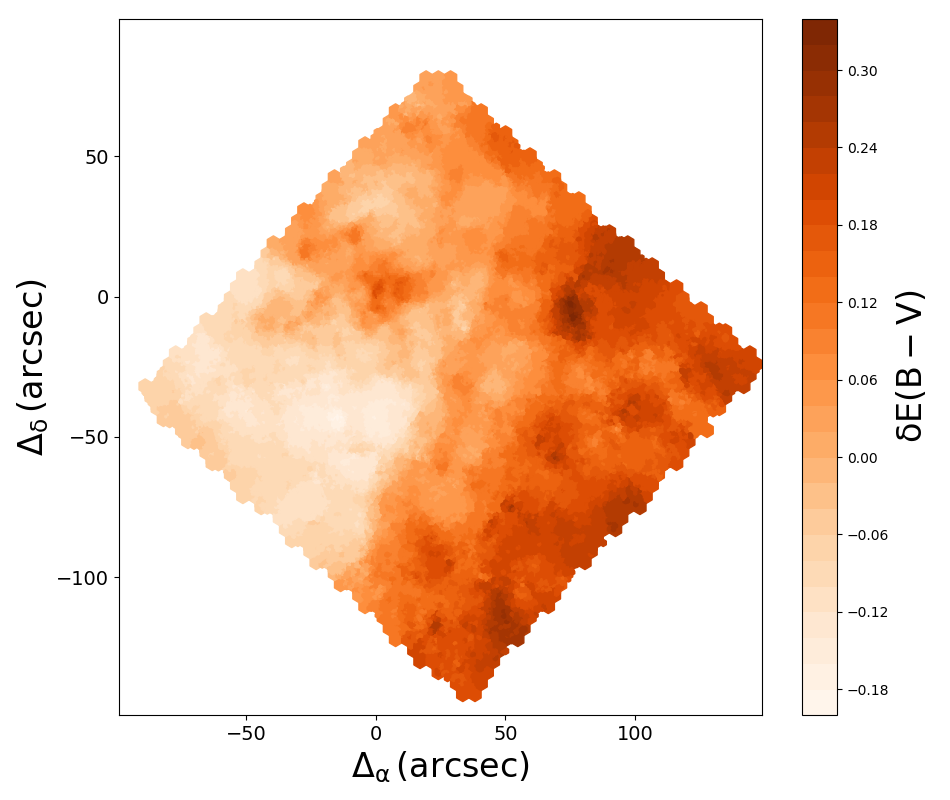}
\caption{Differential reddening map within the observed field of view
  in the direction of NGC 6256. Positions are reported with respect to
  the cluster center. Different colors correspond to different values
  of $\delta E(B-V)$, as reported in the color-bar on the
  right. }
\label{fig:ebvmap}
\end{figure}

At the end of the procedure, we found reddening variations within the surveyed field as large as $\delta E(B-V)=0.51$. The detailed spatial distribution of the reddening variations  is shown in the map plotted in Figure \ref{fig:ebvmap}.  This reddening correction, applied to all the stars in the catalog,
resulted in the CMD shown in right panel of Figure \ref{fig:cmd}. As
can be seen, the evolutionary sequences are now significantly better
outlined: indeed, the sub-giant branch, the red giant branch and an
extended blue horizontal branch can now be clearly appreciated, and
the main sequence is also much thinner. The CMD also reveals that NGC~6256 suffers from a large contamination from field interlopers, that
cannot be removed with the available data-sets because of their
limited time baseline, which make them not suitable to perform proper
motion analysis.

\section{Age, distance and color excess determination}
\label{sec:params}

\subsection{First-guess estimates of distance and color excess}
\label{sec:confronto}
The differential reddening-corrected CMD discussed above has been used
to determine the cluster properties (distance, absolute color excess,
and stellar population age) via isochrone fitting. To this aim, it is
helpful to obtain an independent estimates of both the cluster distance modulus and color excess, to be used as starting point in the isochrone fitting procedure.  As discussed in Section
\ref{sec:intro}, the values for these parameters available in the
literature are still extremely uncertain.

Unfortunately, no RRLyrae stars (that would be suitable to determine
the distance modulus) are known in this cluster. Moreover, the
publicly available parallaxes in the Gaia Data Release 2 (DR2) catalog
\citep{gaia2018} are extremely uncertain for the stars in the
direction of this cluster  and they cannot be used to constrain the cluster distance. The same holds for the absolute color excess: the spatial resolution of the publicly available all-sky maps of interstellar extinction
\citep[e.g.][]{schlafly11} are too poor to get a reliable average
value of this parameter, given the extreme variations on very small
scales shown in Figure \ref{fig:ebvmap}.

\begin{figure}[h!] 
\centering
\includegraphics[scale=0.6]{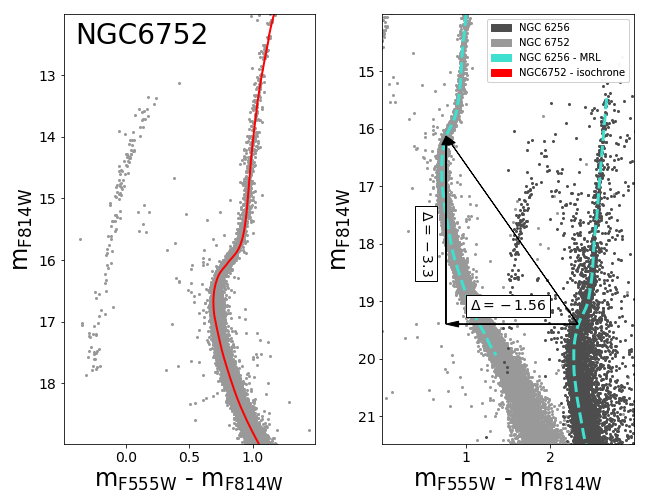}
\caption{{\it Left panel:} $m_{\rm F814W}, (m_{\rm F555W}-m_{\rm
    F814W})$ CMD of NGC~6752. The red line
  is a DSED isochrone with an age of 13 Gyr, reported into the
  observational plane by assuming $E(B-V)=0.04$ and
  $(m-M)_0=13.14$. {\it Right panel:} The dark gray dots show the same
  CMD of NGC~6256 as in Figure \ref{fig:cmd}, while the light gray dots show the same NGC~6752 CMD as in the left panel. The cyan dashed curve
  is the MRL derived in Section \ref{sec:dr}. The adopted shifts in color and
  magnitude are marked with the two black arrows.}
\label{fig:m3}
\end{figure}

We thus estimated the cluster distance modulus and average color excess by comparing the observed CMD with a catalog of stars of the GC NGC~6752, obtained from images acquired under GO 11904 (PI: Kalirai) with the same instrument and combination of filters as for NGC~6256. The data reduction and calibration of these images is exactly the same as the one described in Section~\ref{sec:data}.

NGC~6752 is a $\sim13$ Gyr old system \citep{gratton03,dotter10}, with a metallicity [Fe/H]$=-1.56\pm0.04$ \citep{carretta09} comparable to that of NGC~6256, and very well
constrained distance modulus and color excess: $(m-M)_0=13.14\pm0.06$
and $E(B-V)=0.04 \pm 0.01$ \citep{baumgardt2019,ferraro99}.  For two systems with
approximately the same metallicity and age, it is expected that any
magnitude difference between their MRLs is mainly due to the relative
difference in distance and color excess. The left panel of Figure
\ref{fig:m3} shows that, assuming the metallicity, distance modulus,
and color excess quoted above for NGC~6752, a 13 Gyr isochrone extracted
from the Dartmouth Stellar Evolutionary Database (DSED;
\citealp{dotter2008}) provides a good match of the cluster
CMD.  We then applied color and magnitude shifts to the MRL of NGC~6256 obtained in Section \ref{sec:dr} (cyan dashed line in the right panel of Figure\ref{fig:m3}) to superimpose it onto the MRL of NGC~6752 (created using the same procedure described in Section~\ref{sec:dr}).  To do this, we performed a fit of the two curves in the magnitude range $15<m_{\rm
  F814W}<18$ and chose the best-fit shifts on the basis of the
$\chi^2$ statistics, finding $\Delta(m_{\rm F555W}-m_{\rm
  F814W})=-1.56$ and $\Delta m_{\rm F814W}=-3.3$. Adopting the same extinction coefficients used in Section \ref{sec:dr}, these shifts corresponds to a color excess $E(B-V)=1.18\pm0.05$ and a distance
modulus $(m-M)_0=14.25\pm0.1$ for NGC~6256. We stress that these are only first-guess estimates, just used as starting points for the accurate isochrone fitting of the cluster CMD discussed in the next Section. {In fact, these relative measurements could be biased by the possible differences in the cluster's relative age and in the light-element abundances of the cluster's sub-populations. Moreover, these results could also be biased by the effects of a very different $E(B-V)$ on the observed sequences (see Section~\ref{sec:iso} and Figure~\ref{fig:alambda}).}

\subsection{Isochrone fitting}
\label{sec:iso}
To derive the absolute age of the cluster via isochrone fitting, we
followed a Bayesian procedure similar to that used by \citet[][see also \citealp{correnti2016,kerber2018,kerber2019,cadelano19}]{saracino19}. This approach allows to estimate the stellar system age through a one-to-one comparison between the observed CMD and a set of theoretical models, simultaneously exploring reasonable grids of values for the relevant cluster parameters (not only the age, but also the distance modulus and color excess).

\begin{figure}[t] 
\centering
\includegraphics[scale=0.4]{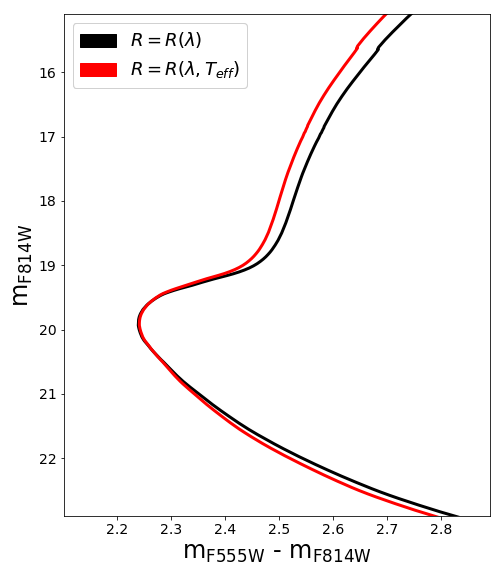}
\caption{The black curve is a DSED isochrone computed for a stellar
  population of 12.5 Gyr and [Fe/H]$=-1.62$, in the same filter
  combination used in this work. It has been put into the
  observational plane by using a distance modulus of 14.25, a color
  excess $E(B-V)= 1.18$, and constant extinction coefficients: $R_{\rm
    F555W}=3.207$ and $R_{\rm F814W}=1.842$. The red curve is the same
  isochrone, but with temperature-dependent extinction coefficients.}
\label{fig:alambda}
\end{figure}

The observed CMD was compared to three different sets of
$\alpha$-enhanced isochrones: the DSED models \citep{dotter2008}, the
Victoria-Regina Isochrone Database (VR, \citealp{vandenberg2014}) and
the BaSTI stellar evolution models \citep{pietrinferni04,
  pietrinferni06}.  For each family of models we assumed
[$\alpha/$Fe]$= +0.4$, which is the typical value for bulge GCs, {and a standard He abundance $Y=0.25$}.  In
the case of the DSED and VR databases, isochrones can be generated at
different metallicities, while for the BaSTI database only a selection
of [Fe/H] values is available. Therefore, in order to perform a proper
comparison between the results obtained from these different models,
we assumed in all cases [Fe/H]$=-1.62$, a value which is extremely
close to that derived through spectroscopy \citep{vasquez18} and that
is available for all the three sets of models.  
The adopted extinction coefficients in the F555W and F814W bands take
into account the dependence of the reddening on the effective
temperature of the stars, following the extinction laws of
\citet{cardelli89}, the equations in \citet{girardi02} and assuming the alpha-enhanced spectral library used in \citet{pietrinferni06}.  This is indeed necessary for performing reliable isochrone fitting in the case
of highly reddened systems, because neglecting such a temperature
dependence makes the red giant branch and the main sequence shallower,
thus increasing the overall color range spanned by the isochrone in
the CMD. This is shown in Figure \ref{fig:alambda}, where we plot, as
an example, the same DSED isochrone (with an age of 12.5 Gyr,
[Fe/H]$=-1.62$ and the first-guess values of distance modulus and
$E(B-V)$ estimated for NGC~6256 in the previous Section) before and
after such a correction: the black line corresponds to the isochrone
obtained by using constant extinction coefficients ($R_{\rm
  F555W}=3.207$ and $R_{\rm F814W}=1.842$), while the red line is
derived by correcting the extinction coefficients for the dependence
on the surface temperature of the stars. As can be seen, the
significant color difference between the two curves demonstrates that
reliable age estimates for heavily reddened systems, such as NGC~6256,
cannot be performed by assuming constant extinction coefficients,
especially when dealing with optical filters (in fact, this effect is
larger for decreasing wavelengths).

To compare the observed CMD of NGC~6256  with the three families of stellar models we applied a Markov Chain Monte Carlo (MCMC) sampling technique. We assumed that the probability of a $n^{th}$ star to belong to an isochrone (corrected for a given reddening and distance) can be expressed in terms of a Gaussian distribution:
\begin{equation}
p_n = \frac{1}{\sqrt{2\pi}\sigma_{n}} \exp{\left[-\frac{1}{2} \frac{d_n^2}{\sigma_{n}^2}\right]} \propto \frac{ \exp{    \left( -\frac{d_n^2}{2\sigma_{n}^2}  \right)   }}{\sigma_{n}}
\end{equation}
where $d_n$ is the minimum distance of the star from the isochrone and $\sigma_{n}$ the uncertainty of such a distance expressed as  $\sigma=\sqrt{\sigma_{col}^2 + \sigma_{mag}^2 + \sigma_{DR}^2}$, where $\sigma_{col}$, $\sigma_{mag}$ are the photometric uncertainties on the color and F814W magnitude, respectively, while $\sigma_{DR}$ are the uncertainties on the differential correction derived as discussed in Section~\ref{sec:dr}.  Therefore, the logarithmic likelihood function of a given isochrone can be expressed as the logarithm of the sum of $p_n$ over all the $n^{th}$ stars:
\begin{equation}
\ln \mathscr{L} = \ln \sum_{n=1}^{N_{stars}} p_n \propto -\frac{1}{2}\sum_{n=1}^{N_{stars}} \left[ \frac{d_n^2}{\sigma_n^2} +\ln(\sigma_{n}^2) \right]
\end{equation}
To sample the posterior probability distribution in the n-dimensional parameter space,  we used the \texttt{emcee} code \citep{foreman19}. Since no age estimates are available in the literature for NGC~6256, we explored a wide range of old ages, uniformly distributed from 10 to 15 Gyr, in steps of 0.25 Gyr.  To put the isochrones into the observational plane, we used values of the color excess and distance modulus following Gaussian
prior distributions peaked at $E(B-V)=1.2\pm0.1$ and $(m-M)_0=14.25\pm0.1$, respectively. In order to minimize the contamination by field interlopers, we extended these calculations only to stars within the cluster half-light radius \citep[$\sim50\arcsec$;][]{harris10}. Moreover, {we restricted the analysis only to stars in the magnitude range $18.5<m_{F814W}<20.5$, a CMD region surrounding the turn-off and therefore sensitive to stellar age variations.}

\begin{figure}[b] 
\centering
\includegraphics[scale=0.42]{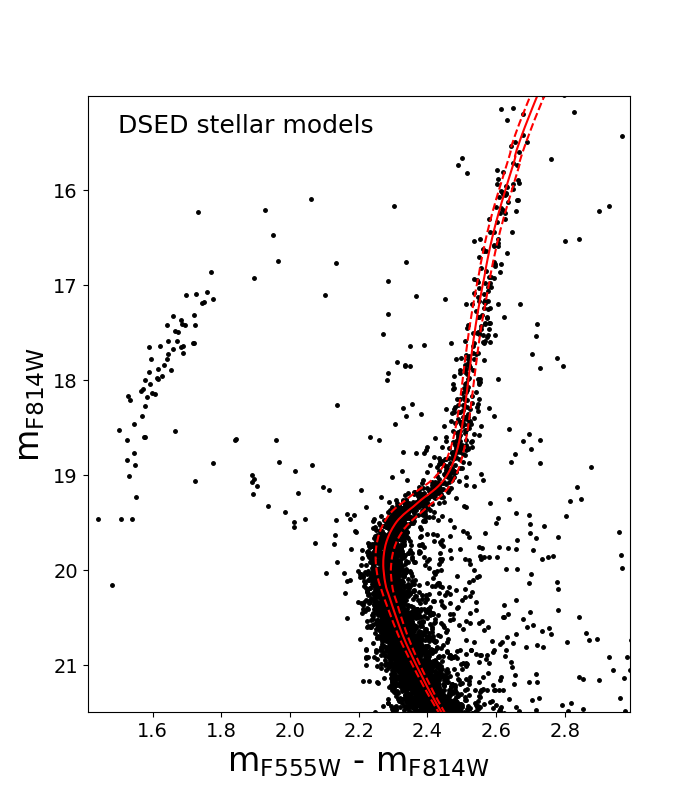}
\includegraphics[scale=0.4]{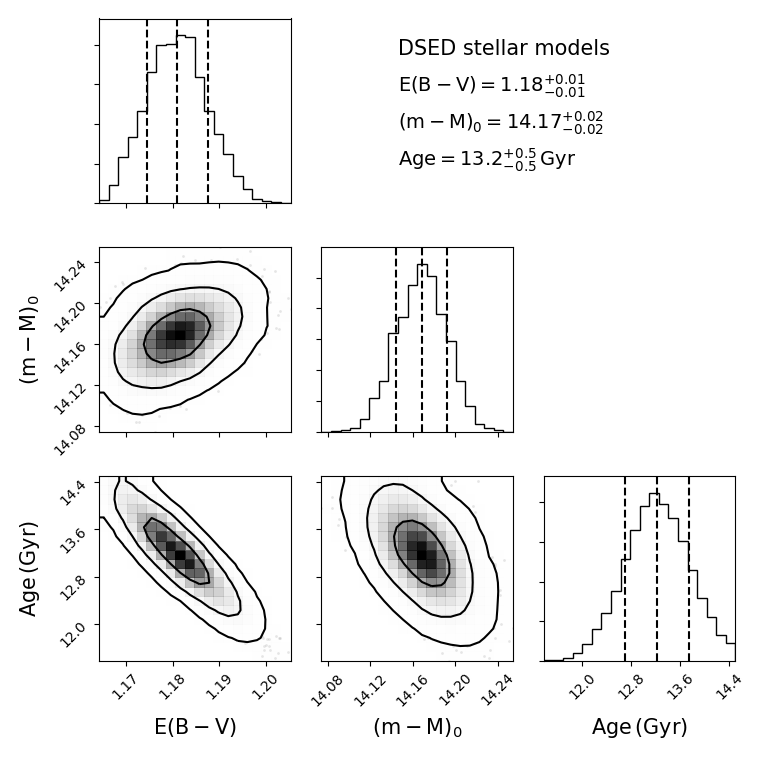}
\caption{ {\it Left panel}: CMD of NGC 6256 with the best-fit DSED isochrone plotted as a red solid line. All the solutions within the $1\sigma$ uncertainties are confined in the region between the two dashed red isochrones. {\it Right panel}: Corner plots
          showing the one- and two-dimensional projections of the
          posterior probability distributions for all the parameters
          derived from the MCMC method applied to the DSED model
          family. The contours correspond to the 1$\sigma$, 2$\sigma$
          and 3$\sigma$ levels. The best-fit parameter values are
          presented in Table \ref{tab:1}.\label{fig:dsediso}}
\end{figure}

\begin{figure}[!ht] 
\centering
\includegraphics[scale=0.42]{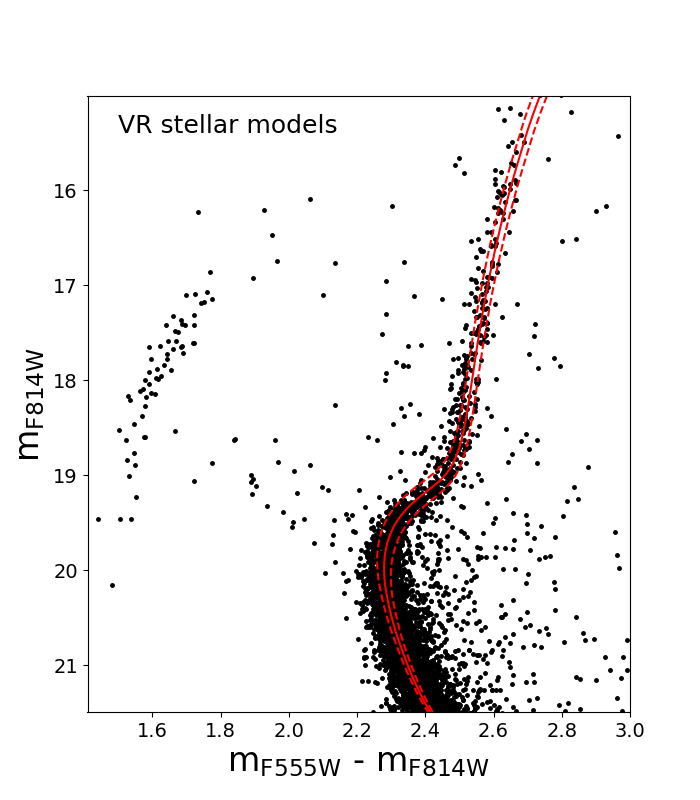}
\includegraphics[scale=0.4]{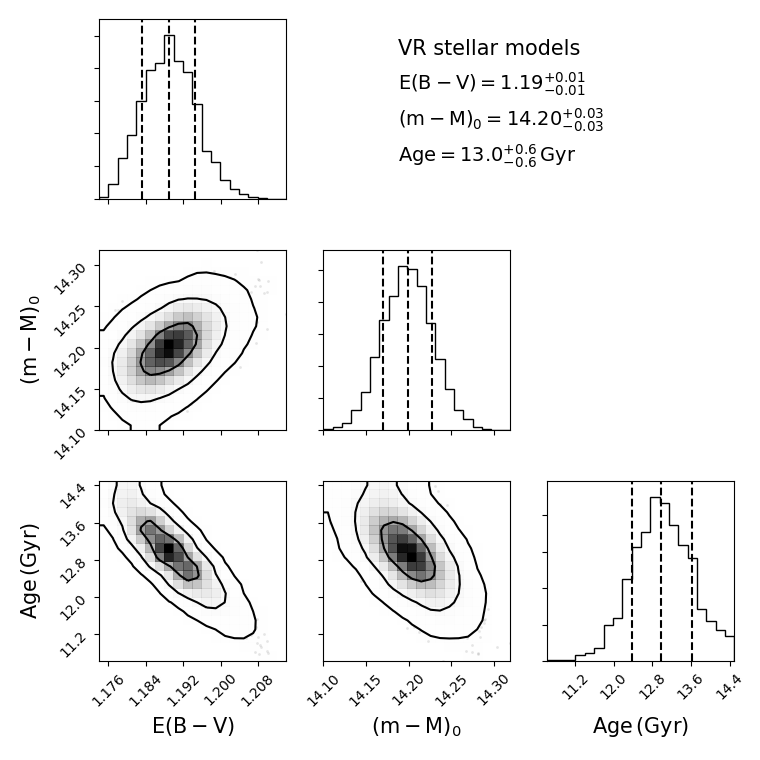}
\includegraphics[scale=0.42]{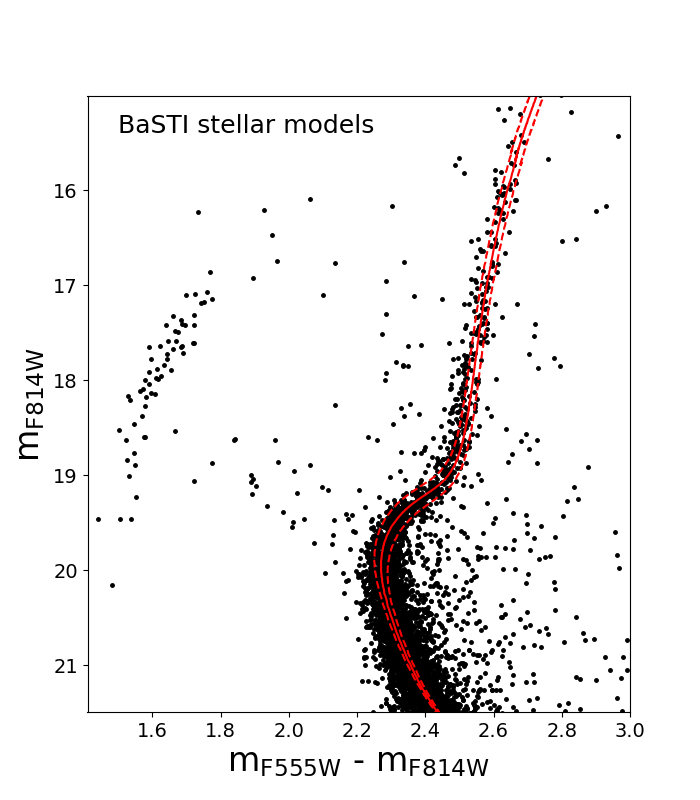}
\includegraphics[scale=0.4]{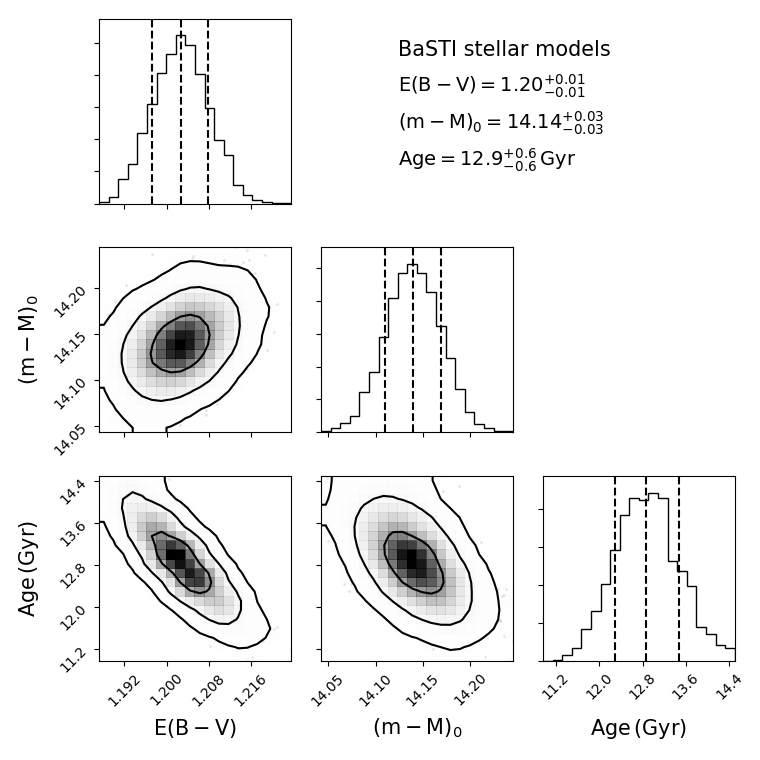}
\caption{As in Figure \ref{fig:dsediso}, but for the VR models (top
  panels) and the BaSTI models (bottom panels). The best-fit parameter
  values are presented in Table \ref{tab:1}.}
\label{fig:vr_basti}
\end{figure}

The results obtained in terms of age, distance modulus and color
excess are shown in Figures \ref{fig:dsediso} and \ref{fig:vr_basti}
for the three adopted sets of theoretical models. In each figure, the
left-hand panel shows the $(m_{\rm F814W}, m_{\rm F555W}-m_{\rm
  F814W})$ CMD  and the best-fit isochrones. In all the cases, the best-fit model very well
reproduces the cluster evolutionary sequences. The
one- and two-dimensional posterior probabilities for all of the
parameter combinations are presented in the right-hand panels as
corner plots. The best-fit values and their uncertainties (based on
the $16^{th}$, $50^{th}$, $84^{th}$ percentiles) obtained for the age,
$E(B-V)$ and $(m-M)_0$ from the three sets of models are also listed in
Table \ref{tab:1}.  Since we are using different
models (adopting slightly different solar abundances, opacities,
reaction rates, efficiency of atomic diffusion etc.), the resulting
best-fit values are somewhat expected to be not exactly the
same. However, they are all mutually consistent within the errors.
NGC~6256 turns out to be a very old cluster, with an age of
13.0 Gyr (obtained as the average of the three derived
values).  The average color excess, $E(B-V)=1.19$, is in good agreement with that quoted by \citealp{valenti07}, and it implies that, within
the field of view covered by our observations, this parameter varies
from a minimum of $\sim1.0$ up to $\sim1.5$.  The obtained distance
modulus corresponds to a distance of $6.8 \pm 0.1$ kpc from the Sun
and turns out to be significantly different from some of the values quoted in literature. Indeed, the cluster results 3.5 kpc closer than reported in \citet{harris10} and 2.3 kpc closer than the value quoted in \citet{valenti07}. 

\begin{table}[ht!]
	\begin{center}
		\caption{Best-fit parameter values for DSED, VR and BaSTI models.}\label{tab:1}
		\begin{tabular}{|c|c|c|c|} 
			\hline
			Model	&   Age    & $E(B-V)$   & $(m-M)_{0}$ \\
				& [Gyr]   &       [mag]    &  [mag]     \\			
			\hline
			DSED   & $13.2\pm0.5$ &  $1.18\pm0.01$ & $14.17\pm0.02$ \\ 
			VR     & $13.0\pm0.6$ &  $1.19\pm0.01$ & $14.20\pm0.03$ \\
			BaSTI  & $12.9\pm0.6$ &  $1.20\pm0.01$ & $14.14\pm0.03$  \\
            \hline
            Average & $13.0\pm0.6$ & $1.19\pm0.01$ & $14.17\pm0.03$ \\  
			\hline
		\end{tabular}
	\end{center}
\end{table} 

As a consistency check, we verified that the results remain basically
unchanged if a uniform prior spanning a large interval of values for
both the color excess and the distance modulus is assumed. As an
additional test, we repeated the isochrone fitting procedure including
the cluster metallicity as a fit parameter. This was however feasible
only for the DSED and the VR models. We allowed the metallicity of the cluster to vary from $-1.1$ dex to $-1.8$ dex, following a normal
distribution with peak value [Fe/H]$= -1.61$ and standard deviation equal to
0.2, as derived by \citet{vasquez18}. The results in terms of age,
color excess and distance modulus are basically unchanged with respect to those quoted in Table \ref{tab:1}, while the best-fit metallicity values are [Fe/H]$=-1.6\pm0.1$ for both the DSED and VR models. They are in excellent agreement with the values derived through spectroscopy.

{Throughout the whole analysis, we assumed that the cluster stellar population has a standard He content $Y = 0.25$. The presence of a He-enhanced sub-population of stars could introduce a broadening of the sequences which should be visible in optical CMDs. We compared the observed width of the evolutionary sequences in the CMD with that derived from a synthetic stellar population having a standard He content. This was created generating stars with F555W and F814W magnitudes randomly extracted from the best-fit BaSTI model convoluted with the observed  error distribution in each filter: $\sigma=\sqrt{\sigma_{mag}^2+\sigma_{DR}^2}$. We found that the width of the synthetic CMD is comparable to the observed one along all the evolutionary sequences, thus confirming that the possible presence of a He-enhanced sub-population cannot be assessed by using the current data-set, as the sequence broadening is dominated by the residuals of the differential reddening corrections. While for a cluster with a mass around $1\times10^5 \ M_{\odot}$, such as NGC~6256 \citep{baumgardt18}, it is expected the presence of sub-populations with a maximum He spread $\delta Y_{max}=0.01-0.02$ \citep{milone18},  we repeated the MCMC analysis using BaSTI models for a stellar population exclusively composed of stars with $Y = 0.3$ ($\delta Y=0.05$). We found that the derived age is consistent within the uncertainties with those derived previously (Table~\ref{tab:1}). Therefore we conclude that, with the available photometry, variations of He abundances expected for this clusters do not have significant impact on the results presented here.}

\section{Proper motion and orbit of NGC 6256}
\label{sec:pm_orbit}

\subsection{The cluster center of gravity}
\label{sec:center}
The gravitational center of the cluster was determined following an
iterative procedure based on the position of resolved stars and
described in \citet[][see also \citealt{montegriffo95,cadelano17a, lanzoni19}]{lanzoni10}. The procedure starts by determining the distance from a first-guess center of all the stars included within a circle of radius $r$
centered on it, and then adopts as new center the average of the star
coordinates. { The procedure is then iteratively repeated: during each iteration the starting center is that derived in the previous iteration and the procedure stops when convergence is reached} (i.e., when the newly-determined center differs by less than $0.01\arcsec$ from the 10 previous determinations). We performed this procedure several times, adopting different values of $r$ and selecting stars in different magnitude ranges, chosen as a compromise between having high enough statistics and avoiding spurious effects due to incompleteness and saturation. In particular, the
radius $r$ was chosen in the range from $20\arcsec$ to $30\arcsec$,
spaced by $5\arcsec$. 
For each radius $r$, we explored different
magnitude ranges, from $m_{\rm F814W}>16.0$ (in order to exclude stars
close to the saturation limit), down to $m_{\rm F814W}=20-21$, in
steps of 0.5 mag. We also excluded from the procedure stars with
color $2<(m_{\rm F555W}-m_{\rm F814W})<3$, in order to minimize the
contamination due to field interlopers. As first-guess center, we used
the value quoted by \citet{harris10}. The final coordinates adopted
for the cluster gravitational center are the mean of the different
values obtained in the different runs, and the uncertainty is their
standard deviation of the measures: $RA=16^{\rm h}59^{\rm
  m}32.668^{\rm s}$ and $Dec=-37^{\circ}7\arcmin15.139\arcsec$, with
an uncertainty of about $0.4\arcsec$. Our newly determined center
differs by $\sim2\arcsec$ from that quoted in \citet{harris10}.

\subsection{The cluster proper motion}
\label{sec:moto}
\begin{figure}[h!] 
\centering
\includegraphics[scale=0.45]{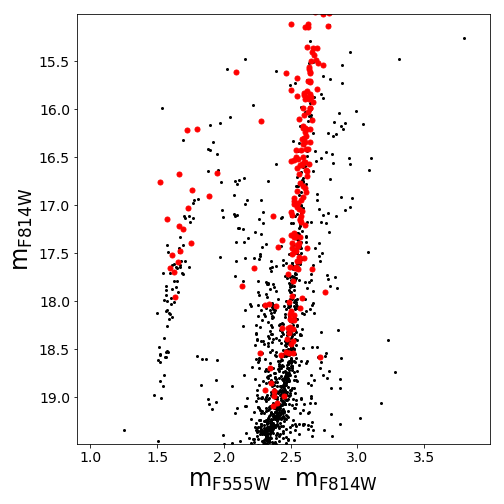}
\includegraphics[scale=0.45]{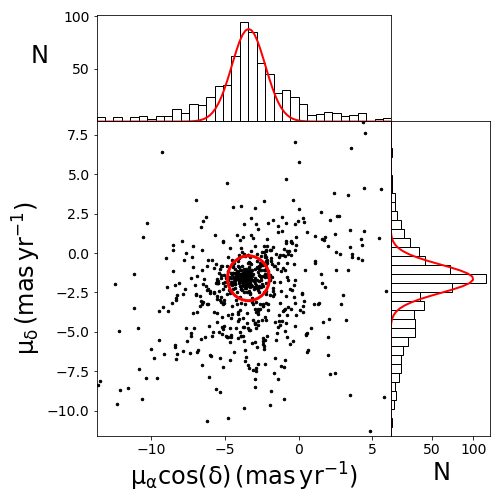}
\caption{{\it Left panel:} Differential reddening-corrected CMD of NGC
  6256 (same as in the right panel of Figure \ref{fig:cmd}), showing
  only the stars in common with the Gaia DR2 catalog. The red dots
  correspond to the stars included within the red circle shown in the
  VPD in the right panel of the figure.  {\it Right panel:} The main panel shows the VPD of the stars in common between our HST catalog and the Gaia DR2 catalog. The upper and side plots show the histograms of the proper motions along RA and Dec, with superimposed a best-fit Gaussian function. The red circle in the main panel is centered on the mean values of the best-fit Gaussian functions  and has a radius of 1.4 mas yr$^{-1}$, equal to the combined $1\sigma$ values. This circle encloses all the stars that have been used to measure
  the cluster absolute PM.}
\label{fig:moti}
\end{figure}

To compute the bulk absolute motion of NGC~6256 on the plane of the
sky, we made use of the absolute proper motions (PMs) of individual
stars available in the Gaia DR2 catalog \citep{gaia2016,
  gaia2018}. First, we selected all the stars in common between our
HST catalog and the Gaia DR2 data set having an absolute PM
measure. Then, we rejected all the objects with poorly measured PM
according to the prescriptions given in \citet{arenou2018}. The CMD
and the vector-point diagram (VPD) of all the stars in common between our observations and the Gaia catalog are shown in Figure~\ref{fig:moti}. We further refined the selection by considering only those stars included within a circle centered on the overdensity of points clearly visible in the VPD, whose center and radius were derived as the mean and standard deviations of the best-fit Gaussians of the histograms shown in the right panel of Figure~\ref{fig:moti} ($\mu_\alpha\cos\delta=3.6\pm1.0$ mas yr$^{-1}$, $\mu_\delta \approx -1.6\pm1.0$ mas yr$^{-1}$). These stars are plotted as red dots in the CMD of Figure \ref{fig:moti}).  The absolute PM of NGC~6526 was finally measured from this fiducial sample of 189 stars, by using the Gaussian Maximum-likelihood method described in \citet[][see also equation 3 of   \citealt{pryor1993}]{walker2006} stars. We found: $\mu_{\alpha}\cos(\delta)=-3.7\pm0.2$ mas yr$^{-1}$ and
$\mu_{\delta}=-1.6\pm0.2$ mas yr$^{-1}$, in the J2000.0 system. These
values are consistent with those determined by \citet{vasiliev2019}
and \citet{baumgardt2019}, which are both based on Gaia DR2 PMs as
well. 

\subsection{The cluster orbit}
\label{sec:orbit}
\begin{figure}[h] 
\centering
\includegraphics[scale=0.4]{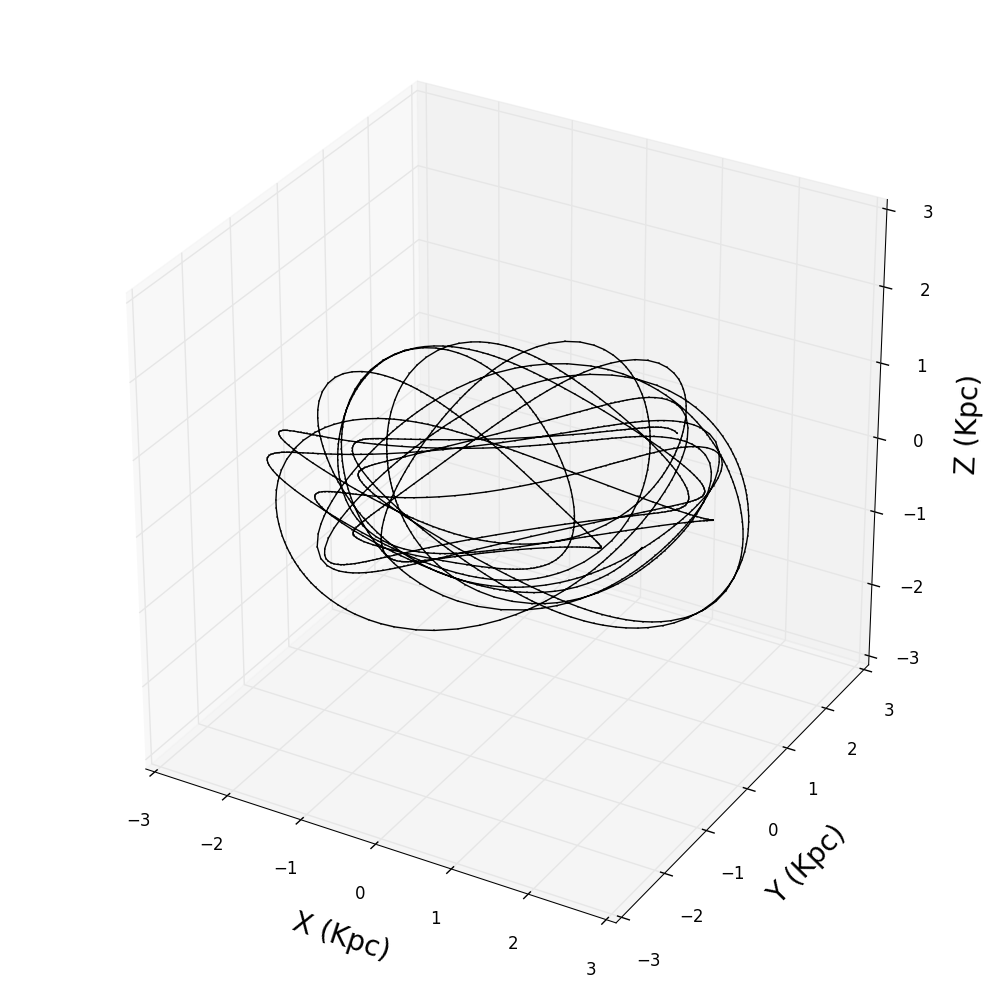}
\caption{Simulated positions occupied by NGC 6256 in the last Gyr
  during its orbit within the Galaxy. The 3D axes cover a region
  smaller than the size of the Galactic bulge, thus confirming that
  the orbit of this cluster is well confined within it.}
\label{fig:orbit}
\end{figure}

The cluster absolute PM thus derived, combined with the radial
systemic velocity from \citet[][$v_r =-103.4 \pm 0.5$ km
  s$^{-1}$]{vasquez18}, was used to back-integrate the cluster orbit
within the Milky Way potential well. To this purpose, we used {\rm
  GravPot16}\footnote{\url{https://gravpot.utinam.cnrs.fr/}}
\citep{fernandez17}, a package that generates stellar orbits in a
semi-analytic, steady-state and three-dimensional Galaxy model, based
on the gravitational potential derived from the Besan\c{c}on Galactic
Model \citep{robin03} and including also a prolate bar
\citep{robin12}. This tool has been recently used to compute the
orbits of several Galactic star clusters \citep{gaia2018, libralato18,
  bellazzini19}. The Galaxy parameters (such as the bar properties,
Sun distance from the Galactic center, etc.) were set to the same
standard values adopted by \cite{gaia2018}.  Figure \ref{fig:orbit}
shows the resulting orbit of the cluster during the last Gyr.  We
verified that nothing significantly changes if the cluster initial
conditions are randomly varied within the uncertainty ranges of its
position and velocity.  It can be seen that NGC~6256 is in a
low-eccentricity orbit ($e\sim0.11$), strongly confined within the
Galactic bulge. In fact, during its $\sim50$ Myr orbital period, it
reaches a minimum and maximum distance from the Galactic center of
about 0.8 kpc and 2.9 kpc, respectively. This confirms that, despite
its relatively low metallicity, this system is not a halo intruder
crossing the bulge during a fraction of its orbit (as suggested for
other bulge GCs, like NGC~6681, Terzan 10 and Djorgovski~1,
\citealp{massari13,ortolani19a}), but it is a GC genuinely belonging to the bulge
population. As a consistency check, we compared our results with those obtained using different orbit integrators \citep[see, e.g.][]{cadelano17a} that adopt different shapes of the Galactic potential. We found that all
the results are consistent within the uncertainties. These results are
also qualitatively in agreement with those reported by
\citet{baumgardt2019}. It is worth stressing that the simulated cluster orbit cannot be used to firmly asses if the cluster birthplace was indeed the Galactic bulge, since a
back-integration of the orbit for a time as long as the cluster age
should take into account the (basically unknown) variation of the
Galactic potential as a function of time.  

\begin{figure}[h] 
\centering
\includegraphics[scale=0.45]{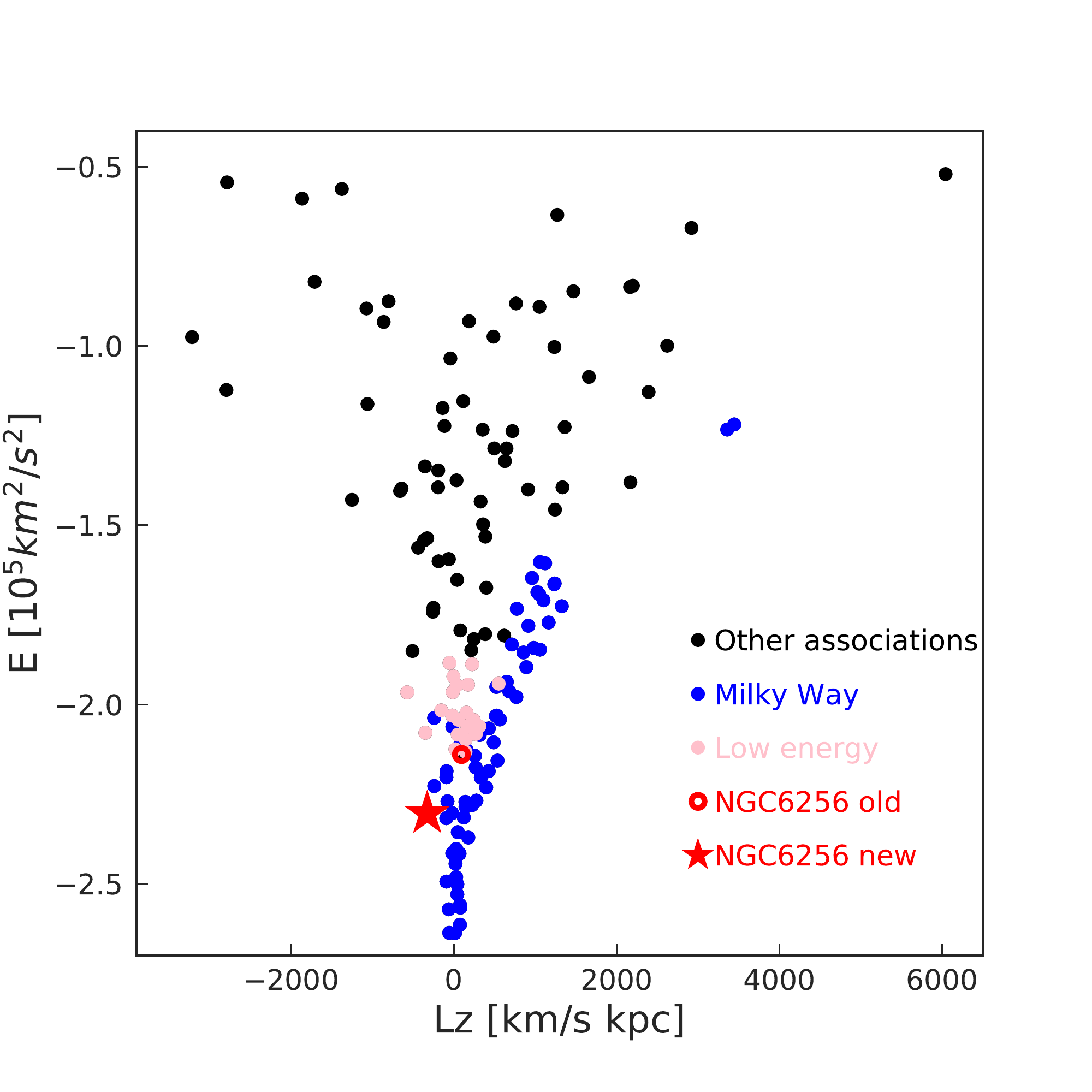}
\caption{Integrals of motion for the sample of Galactic GCs analyzed in \citet{massari19}. Blue points are clusters belonging to the Galaxy in situ population, pink points are those associated with an accreted low-energy component while black points are clusters with different associations. The position of NGC~6256 previously computed by \citet{massari19} is reported with a red open circle while its updated position is shown with a red star.}
\label{fig:iom}
\end{figure}

\citet{massari19} calculated the integrals of motion for a large sample of Galactic GCs to infer their birthplaces (i.e. to infer which clusters belong to the main Galactic population and which ones have been likely accreted during merger events that the Galaxy experienced in the past). To this aim they used the cluster current positions from \citet{harris10} and kinematics from the Gaia DR2 catalog. They found that NGC~6256 is unlikely to belong to the Galaxy in situ population and it is most likely associated with an accreted low-energy component (see Figure~3 in \citealp{massari19}). However, their result is based on a distance value significantly different from the one computed here, while the cluster kinematic is in agreement with the results derived here and by \citet{vasquez18}. We therefore repeated the same procedure followed by \citet{massari19}  using the updated cluster position and found that NGC~6256 has a orbital energy of  $E\sim -2.3\times10^5$ km$^{2}$/s$^{2}$ and an angular momentum of $L_z\sim -300$ km/s kpc. These updated values place NGC~6256 in a region of the integrals of motion space occupied by systems that genuinely belong to the population of bulge clusters  formed in situ (Figure~\ref{fig:iom}). 

These dynamical arguments strongly suggest that NGC~6256 is an old and low-metallicity GCs that was formed in the bulge during the very early epoch of the Galaxy assembly.

\section{Summary and conclusions}
\label{sec:conc}
We exploited, for the first time, deep and high resolution optical
observations of the bulge globular cluster NGC~6256 to derive the main
properties of its stellar population. The analysis revealed a
population affected by high and differential reddening, which causes a
severe blurring of all the cluster evolutionary sequences. We
corrected the CMD position of each observed star for this effect and
created a differential reddening map that reveals $E(B-V)$ variations
up to $0.5$ mag across the $160\arcsec \times 160\arcsec$ field of
view covered by the observations. In the differential
reddening-corrected CMD, all the evolutionary sequences are nicely
defined, and the main sequence is sampled down to 4 magnitudes below
the turn-off. Our photometry confirmed that NGC~6256 is an outsider
system with respect to most of the bulge GCs, since its stellar
population is characterized by an extended and blue horizontal branch,
suggesting a relatively low metal content, as confirmed by spectroscopic studies. These rare, relatively metal-poor GCs in the bulge
are expected to be the oldest relics of the Galaxy assembly process
and we thus performed isochrone fitting to determine the cluster
stellar population age, along with other properties such as its
distance and average color excess. It turned out that NGC~6256 is one
of the oldest clusters known to date in the Galactic bulge. Indeed, the comparison with
different sets of isochrones revealed a stellar population having an average age of
$13.0$ Gyr with a typical uncertainty around $0.5$ Gyr. This result is particularly worth of attention:
although bulge GCs are known to be 12 Gyr-old on average, improved age estimates are recently suggesting that metal-rich clusters are
slightly younger (see the case of NGC 6528 in \citealt{lagioia2014,
  calamida2014}), while metal-poor GCs are preferentially older than
this limit (NGC 6558, NGC 6522 and HP1; \citealt{barbuy2007,
  kerber2018, kerber2019}). The cluster age here obtained for NGC~6256
fits within such a scenario. We also found that NGC~6256 is affected
by a large absolute color excess $E(B-V)=1.19$ and that its
absolute distance modulus implies a distance of 6.8 kpc from the Sun.
We finally determined the gravitational center and the absolute proper
motion of the cluster, to then derive its orbit within a Galaxy-like
potential. We found that NGC~6256 is moving in a low-eccentricity
orbit entirely confined within the bulge, thus confirming that it is
not a halo intruder crossing the bulge during its motion, but a
genuine member of the Galactic bulge. Moreover, we computed the cluster integrals of motion and found that the cluster binding energy and angular momentum are compatible with those expected for a cluster purely belonging to the in situ Galactic bulge population.
All together, these pieces of evidence indicate that NGC~6256 is one of the oldest systems formed within the bulge during the first stages of Galaxy assembly.

\acknowledgments
{The authors warmly thank the anonymous referee for his/her careful reading of the manuscript.} This paper is part of the project Cosmic-Lab (``Globular Clusters as Cosmic Laboratories'') at the Physics and Astronomy Department of the Bologna University (see the web page: \url{http://www.cosmic-lab.eu/Cosmic-Lab/Home.html}). The research is funded by the project Dark-on-Light  granted by MIUR through PRIN2017 contract (PI:Ferraro).\\
Based on observations with the NASA/ESA Hubble Space Telescope , obtained at the Space Telescope Science Institute, which is operated by AURA, Inc., under NASA contract NAS 5-26555. \\
This work has made use of  data from the European Space Agency (ESA) mission Gaia (\url{https://www.cosmos.esa.int/gaia}), processed by the Gaia Data Processing and Analysis Consortium (DPAC, \url{https://www.cosmos.esa.int/web/gaia/dpac/consortium}).

\vspace{5mm} \facilities{HST(WFC3)} \software{DAOPHOT
  \citep{stetson87}, {\rm emcee} \citep{foreman19}, {\rm corner.py}
  \citep{foreman16}}, {\rm GravPot16} \citep{fernandez17}

\end{document}